\documentclass[preprint,showpacs,preprintnumbers,amsmath,amssymb]{revtex4}
\usepackage{graphicx}
\usepackage{dcolumn}
\usepackage{bm}
\usepackage{graphicx}
\begin{document}
\title{Superconductivity by Kinetic Energy Saving ?}
\author{D. van der Marel, H.J.A. Molegraaf, C. Presura, I. Santoso}
\address{ Material Science Center, University of Groningen, 9747 AG
Groningen, The Netherlands}
\begin{abstract}
A brief introduction is given in the generic microscopic framework
of superconductivity. The consequences for the temperature
dependence of the kinetic energy, and the correlation energy are
discussed for two cases: The BCS scenario and the non-Fermi liquid
scenario. A quantitative comparison is made between the
BCS-prediction for d-wave pairing in a band with nearest neighbor
and next-nearest neighbor hoppping and the experimental specific
heat and the optical intraband spectral weight along the plane. We
show that the BCS-prediction produces the wrong sign for the kink
at $T_c$ of the intraband spectral weight, even though the model
calculation agrees well with the specific heat.
\end{abstract}
\maketitle
\section{Model independent properties of the superconducting state}
\subsection{Internal energy}
\begin{figure}[t]
 \centerline{\includegraphics[width=6cm]{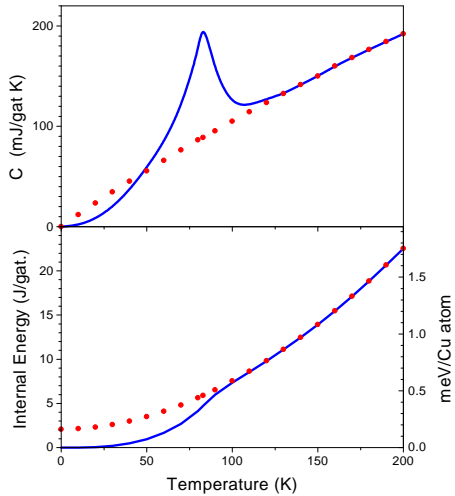} }
 \caption{Experimental internal energy and the specific heat
 obtained by Loram {\em et al}\cite{loram01}. The extrapolations are obtained
 following the procedure of Ref.~\cite{marel02}}
 \label{loramint}
\end{figure}

When we cool down a superconductor below the critical temperature,
the material enters a qualitatively different state of matter,
manifested by quantum coherence over macroscopic distances.
Because the critical temperature represents a special point in the
evolution of the internal energy versus temperature, the internal
energy departs from the temperature dependence seen in the normal
state when superconductivity occurs. Because for $T < T_c$ the
superconducting state is the stable equilibrium state, the
internal energy in equilibrium at $T=0$ is an absolute minimum.
Hence cooling down from above the phase transition one would
expect a drop in the internal energy when at the critical
temperature. This drop of internal energy stabilizes the
superconducting phase among all alternative states of matter. An
experimental example of this well known behavior is displayed in
Fig.~\ref{loramint}, where the internal energy was calculated from
the electronic specific heat data\cite{loram01} according to the
relation $E_{int}(T)=\int_0^T c(T^{\prime})dT^{\prime}$.
The
broadened appearance of the phase transition suggests that the
superconducting correlations disappear rather gradually when the
temperature is increased above the phase transition, an effect
which remains noticeable in the specific heat graph up to 125 K.
\\
Understanding the mechanism of superconductivity means to
understand what stabilizes the internal energy of the
superconducting state. In BCS theory superconductivity arises as a
result of a net attractive interaction between the quasi-particles
of the normal state. Note, that implicitly this approach is firmly
rooted in the paradigm of a Fermi-liquid type normal state. On the
other hand, the school based on Anderson's original
work\cite{anderson87} asserts, that a strong on-site {\em
repulsive} interaction can also give rise to high $T_c$
superconductivity. The latter models typically require that the
material is not a Fermi liquid when superconductivity is muted,
either by raising the temperature or by other means.
\subsection{Pair-correlations}
Without loss of generality, {\em i.e.} independent of the details
of the mechanism which leads to superconductivity, it is possible
to provide a microscopic definition of the superconducting state.
For this purpose let us consider the correlation function defined
as
\begin{equation}\label{GRr}
 G(r,R_1,R_2)=\left<\psi^{\dagger}_{\uparrow}(R_1+r)\psi^{\dagger}_{\downarrow}(R_1)
 \psi_{\downarrow}(R_2)\psi_{\uparrow}(R_2+r)\right>
\end{equation}
where $\psi^{\dagger}_{\sigma}(R_j)$ is a single electron creation
operator. This function defines two types of correlation: (i) the
electron pair-correlation as a function of the relative coordinate
$\vec{r}$, and (ii)v the correlation between two pairs located at
center of mass coordinates $\vec{R}_1$ and $\vec{R}_2$. In the
normal state there is no correlation of the phase of
$G(r,R_1,R_2)$ over long distances $|R_1-R_2|$ due to the finite
mean free path of the electrons. As a result the integral over the
center of mass coordinates of the correlation function
\begin{equation}\label{gr}
    g(r)=\frac{1}{V^2} \int d^3R_1 \int d^3R_1 G(r,R_1,R_2)
\end{equation}
averages to zero in the normal state. In contrast the
superconducting state is characterized by long range phase
coherence of the center of mass coordinates, implying (among other
things) that the correlation function averaged over all center of
mass coordinates, $g(r)$, is a finite number.
\section{Internal energy and its decomposition using the BCS model}
\subsection{BCS: Correlation function, and correlation energy for d-wave pairing}
\begin{figure}[t]
  \centerline{\includegraphics[width=6cm]{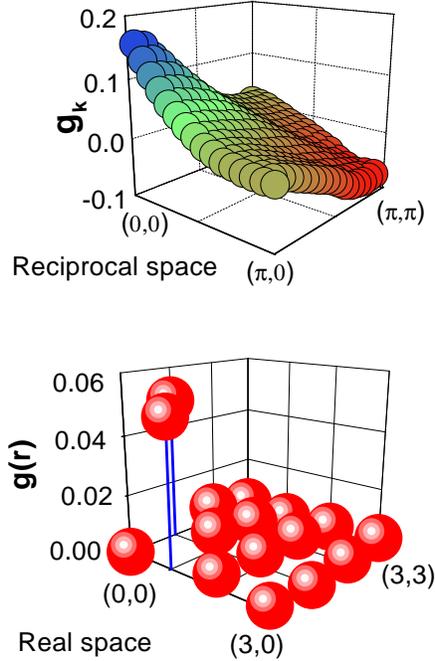}}
  \caption{\protect The  k-space (top panel) and coordinate space (bottom panel)
  representation of the superconductivity induced change of
  pair-correlation function for d-wave symmetry (bottom panel).
  Parameters: $\Delta/W = 0.2$, $\omega_D/W=0.2$. Doping level:
  x = 0.25}
  \label{dgkr}
\end{figure}

In the weak coupling scenario of BCS theory the electrons have an
effective attractive interaction, as a result of which they tend
to form pairs. For the purpose of the present discussion we will
assume that the interaction is of the form
\begin{equation}\label{Hint}
    H^i=V(r-r')\hat{n}(r)\hat{n}(r')
\end{equation}
where $\hat{n}(r)$ is the electron density operator. The
interaction energy in the superconducting state becomes lower than
in the normal state, due to the fact that the effective attractive
interaction favors a state with enhanced pair-correlations. The
value of the interaction energy of the superconducting state,
relative to the normal state is
\begin{equation}\label{Hi}
   \left<H^i\right>_s-\left<H^i\right>_n
    =\int d^3r g(r) V(r)
    =\sum_k g_k V_k
\end{equation}
where $g_k$ and $V_k$ are the Fourier transforms of $g(r)$ and
$V(r)$ respectively. Using the Bogoliubov transformation the
correlation function can be expressed\cite{particle-hole} in terms
of the gap-function $\Delta_k$ and the single particle energies
$E_k=\{(\epsilon_k-\mu)^2+\Delta_k^2\}^{1/2}$.
\begin{equation}\label{gk}
    g_k=\sum_q\frac{\Delta_{q+k}\Delta^*_q}{4E_{q+k}E_q}
\end{equation}
This corresponds to the conversion of a pair $(q,-q)$ to a pair
with quantum numbers $(q+k,-q-k)$. In the expression for the
correlation energy, Eq.~\ref{Hi}, the transferred momentum $k$ is
carried by interaction kernel $V_k$.
\\
Starting from a model expression for the single electron
energy-momentum dispersion $\epsilon_k$, and the gap-function
$\Delta_k$, it is a straightforward numerical exercise to
calculate the summations in Eq.~\ref{gk}. Adopting the nearest
neighbor tight-binding model with a d-wave gap, and adopting the
ratio $\Delta(\pi,0)/W=0.2$, where $W$ is the bandwidth, the
correlation function $g_k$ can be easily calculated, and the
result is shown in Fig.~\ref{dgkr}.
We see from this graph that a negative value of
$\left<H^i\right>_s-\left<H^i\right>_n$ requires either (i)
$V_k<0$ for $k$ in the neighborhood of the origin, or (ii) $V_k>0$
for $k$ in the vicinity of $(\pi,\pi)$. The corresponding
representation in real space, $g(r)$, shown in Fig.~\ref{dgkr},
illustrates that the dominant correlation of the d-wave
superconducting state is of pairs where the two electrons occupy a
nearest neighboring site, while the on-site amplitude is zero.
\\
Combining the information of Fig.~\ref{dgkr} with Eq.~\ref{gr}, it
is clear that the strongest saving of correlation energy is
expected if the electrons interact with an interaction of the form
$V(r_1,r_2)=V_0 \sum_a \delta(\vec{r}_1-\vec{r}_2+\vec{a})$ where
the vector $a$ runs over nearest neighbor sites, and $V_0$ is a
negative (meaning attractive) interaction between electrons on
nearest neighbor sites.
\subsection{BCS: The gap-equation, specific heat and internal energy}
\begin{figure}[t]
 \centerline{\includegraphics[width=6cm]{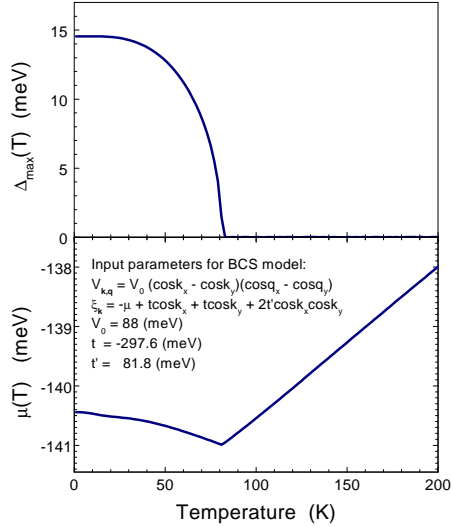}}
 \caption{BCS prediction of the d-wave gap-function and the
 chemical potential.}
 \label{delmut}
\end{figure}

\begin{figure}[t]
 \centerline{\includegraphics[width=6cm]{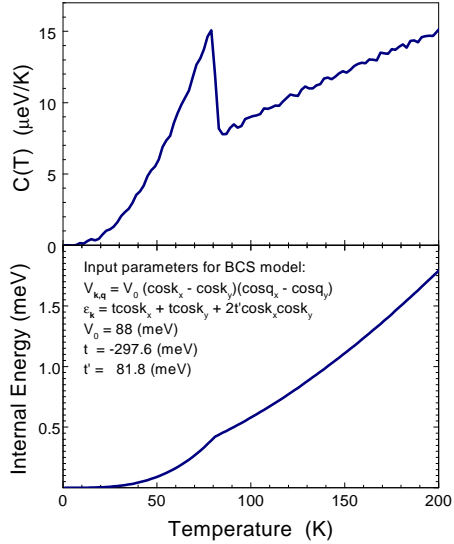} }
 \caption{BCS prediction of the internal energy and the specific
 heat.}
 \label{einternal}
\end{figure}

To illustrate the predictions of BCS theory for the temperature
dependence of the correlation energy and the kinetic energy, we
start by solving the gap equation
\begin{equation}\label{delta(T)}
   \Delta_k=\sum_q\frac{V_{k-q}\Delta_q}{2E_q}\tanh\left(\frac{E_k}{2k_BT}\right)
\end{equation}
which must solved together with the constraint that the average
number of  particles is temperature independent. This requires
that the chemical potential must be adjusted to keep the thermal
average of $\sum_k \left<\hat{n}_k\right>=N$ at a constant value.
In the numerical examples of this paper we have adopted $N=0.85$
corresponding to a hole doping of 0.15. This solution of the gap
equation is shown in Fig.~\ref{delmut}.
Notice, that the temperature dependence of the chemical potential
follows closely the experimental observations reported for Y123 in
Ref. \cite{rietveld92}
\\
To check that the band-parameters used here are reasonable, we
display in Fig.~\ref{einternal} the corresponding prediction for
the specific heat and the internal energy, using the same
parameters as for Fig. \ref{delmut}. If we compare this to
Fig.~\ref{loramint}, we see that the band-parameters adopted here
quantitatively reproduce the observed specific heat. Hence the
present set of band-parameters ($t$,$t^{\prime}$) and the coupling
parameter $V_0$ represent the best phenomenological choice for
quantitative testing of the BCS-model.
\subsection{BCS: Temperature dependence of the correlation energy and the kinetic energy}
\begin{figure}[t]
 \centerline{\includegraphics[width=6cm]{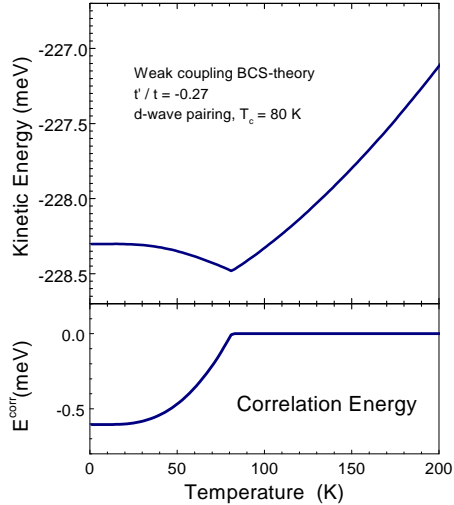} }
 \caption{BCS prediction of the kinetic energy and the correlation
 energy.}
 \label{energyt}
\end{figure}

The BCS prediction for the temperature dependence of the average
interaction energy follows from Eq.~\ref{Hi}. We then use the BCS
variational wave-function for the statistical average of
Eq.~\ref{gk}, resulting in
\begin{equation}\label{Hi(T)}
   \left<H^i\right>_s-\left<H^i\right>_n
    =-\sum_k \frac{|\Delta_k|^2}{2E_k}\tanh\left(\frac{E_k}{2k_BT}\right)
\end{equation}
Simultaneously there is an increase of the 'kinetic' energy
\begin{equation}\label{Ekin(T)}
   \left<H^{kin}\right>
    =\sum_k \epsilon_k\left\{
     1-\frac{\epsilon_k-\mu}{E_k}\tanh\left(\frac{E_k}{2k_BT}\right)
                  \right\}
\end{equation}
In Fig.~\ref{energyt} this is displayed, using the same parameters
as for Fig. \ref{delmut}.
\section{Relationship between intra-band spectral weight and kinetic energy}
\begin{figure}[t]
 \centerline{\includegraphics[width=6cm]{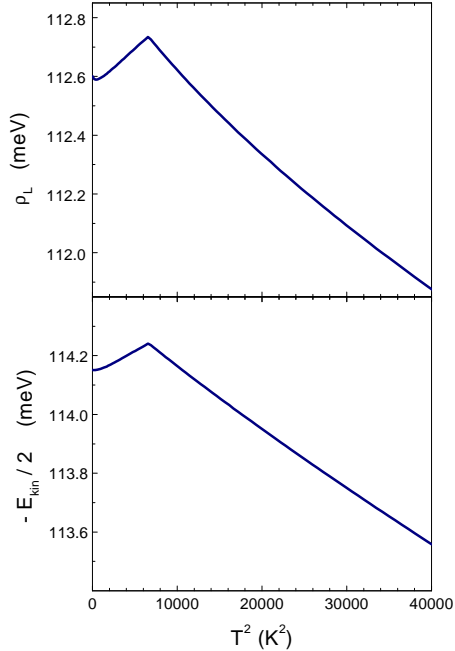} }
 \caption{BCS prediction of the spectral weight function.}
 \label{swth}
\end{figure}

A measure of the kinetic energy is provided by the following
relation\cite{norman02,kubo57,drew97}
\begin{equation}
       \int_{-\Omega}^{\Omega}d\omega\mbox{Re}\sigma(\omega)d\omega
       = \pi \frac{e^2}{\hbar^2V}  \sum_{k,\sigma}
        \langle \hat{n}_{k,\sigma}\rangle
      \frac{\partial^2\epsilon(\vec{k})}{ \partial k^2}
 \label{fsum2}
\end{equation}
where the high frequency limit indicates that the integral should
include only the intra-valence band transitions, and the
condensate peak at $\omega=0$ if the material is a superconductor.
The integral over negative and positive frequencies (note that
$\sigma(\omega)=\sigma^*(-\omega)$) avoids ambiguity about the way
the spectral weight in the condensate peak should be counted. If
the band structure is described by a nearest neighbor
tight-binding model, Eq.~\ref{fsum2} leads to the simple relation
\begin{equation}
 \begin{array}{l}
     \rho_L \equiv \frac{\hbar^2}{a^2 \pi e^2}\int_{-\Omega}^{\Omega}
     \mbox{Re}\sigma(\omega)d\omega = \langle - H_{kin} \rangle
 \end{array}
 \label{fsum4}
\end{equation}
Hence in the nearest neighbor tight-binding limit the {\em
partial} f-sum provides the {\em kinetic energy} contribution,
which depends both on the number of particles and the hopping
parameter $t$\cite{maldague77,carmelo88,basov99,chakravarty99}.
However, if the band-structure has both nearest neighbor hopping
and next nearest neighbor hopping, Eq.~\ref{fsum4} is not an exact
relation, and instead Eq.~\ref{fsum2} should be compared directly
to the experiments. In Fig.~\ref{swth} we compare the spectral
weight, calculated directly using Eq.~\ref{fsum2} to the result of
Eq.~\ref{fsum4}, using the same parameters as for Fig.
\ref{delmut}.
Note that the kinetic energy has to be divided by a factor two, as
we are interested in the projection along one of the two axes in
the ab-plane, which can be compared directly to the experimental
value of $\rho_L$. From Fig.~\ref{swth} we can conclude, that the
effect of including $t^{\prime}$ in the calculation is rather
small, and it is still OK to identify $\rho_L$ with the kinetic
energy apart from a minus sign.
\section{Experimental determination of the intra-band spectral weight}
\begin{figure}[t]
 \centerline{\includegraphics[width=6cm]{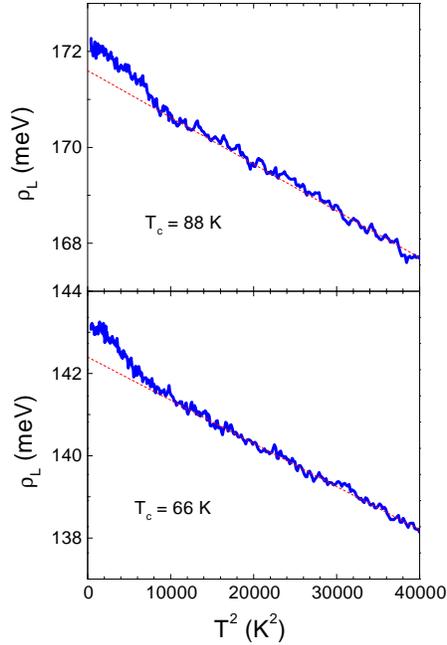} }
 \caption{Experimental values of the ab-plane spectral weight
 function, taken from Ref.~\cite{molegraaf02}}
 \label{swexp}
\end{figure}

In two recent experimental papers measurements of $\rho_L$ in
Bi2212 have been reported \cite{molegraaf02,santander02}. The
values of the kinetic energy change in the superconducting state
were in quantitative agreement with each other, and both papers
arrived at the same conclusion: Contrary to the BCS prediction,
the kinetic energy of the superconducting state is {\em lower}
than in the normal state (taking into account a correction for the
temperature trends of the normal state). In Fig.~\ref{swexp} the
data of Ref. \cite{molegraaf02} have been reproduced. Comparing
this with the BCS prediction clearly demonstrates the large
qualitative discrepancy between theory and experiment. Clearly the
type of mechanism assumed in BCS theory is not at work here.
\section{Implications of the experimental data}
The trend seen in the experimental data has been predicted by
Hirsch in 1992\cite{hirsch92a,hirsch92b,hirsch98}. The model
assumption made by Hirsch was, that the hopping probability of a
single hole between two sites becomes larger if one of the two
sites is already occupied by a hole. Although this model provides
good qualitative agreement with the optical experiments, it has
one serious deficiency: It also predicts s-wave symmetry for the
order parameter, in sharp contrast to a large body of experimental
data which show that the superconducting gap in the cuprates has
d-wave symmetry.
\\
In a recent set of calculations based on the Hubbard model,
Jarrell et al\cite{jarrell03} obtained a similar effect as seen in
our experiments, both for underdoped and optimally doped samples.
Crudely speaking the mechanism is believed due to the frustrated
motion of single carriers in a background with short-range
(RVB-type) spin-correlations, which is released once pairs are
formed.
\section{conclusions}
We have made a quantitative comparison between the BCS-prediction
for d-wave pairing in a band with nearest neighbor and
next-nearest neighbor hoppping and various experiments, in
particular specific heat and measurements of the optical ab-plane
sumrule. We have shown that the BCS-prediction produces the wrong
sign for the kink at $T_c$ of the ab-plane intraband spectral
weight, while the model calculation is in good agreement with the
experimental specific heat data.
\begin{acknowledgements}
DvdM gratefully acknowledges M. Norman, N. Bontemps, J. Hirsch,
and M. Jarrell for stimulating discussions, and J. W. Loram for
making his data files of the specific heat of Bi2212 available.
\end{acknowledgements}

\begin{thebibliography}{99}
\bibitem{particle-hole}
        There is an additional term, due to the fact that the chemical potential
        of the superconducting state is slightly shifted relative to the normal
        state\cite{marel90,marel92}: $0.25\sum_p|v_{p+k}v_p|^2$. In the
        numerical examples in this paper this additional term has been taken
        into account. Since the effect is small, we will not discuss it further here.
\end{thebibliography}
\end{document}